# Electric Field Resolved Image Formation in a Widefield Optical Microscope

Arjun Ashoka[1*], Juhwan Lim[1], Akshay Rao[1], Dmitry A. Zimin[1,2*]

[1]Cavendish Laboratory, University of Cambridge, J.J. Thomson Avenue, Cambridge, CB3 0HE, United Kingdom

[2]Laboratory of Physical Chemistry, ETH Zürich, Vladimir-Prelog-Weg 2, 8049, Zürich, Switzerland

Correspondence: aa2066@cam.ac.uk, dzimin@ethz.ch

**Abstract:** Visualizing the spatiotemporal evolution of the electric field of light is fundamental to optics, from designing photonic devices to developing next-generation microscopes. However, we lack the experimental tools to directly access the electric field of light in the sample plane of an optical microscope. Here, we introduce an all-optical imaging modality that resolves the electric field of light in the plane of a traditional widefield transmission optical microscope with 100-attosecond temporal and 200-nanometer spatial resolution. With this we demonstrate the delayed buildup of scattering contrast and pulse broadening through and around a thick MoTe$_2$ flake – dynamics inaccessible via standard simulations. We showcase our technique's versatility by additionally resolving the full in-plane vector electric field lines during photoexcitation as the optical pulse propagates through and around the MoTe$_2$ flake.

**One-Sentence Summary:** An all-optical ripple-tank.

Since the discovery of Maxwell's equations, our ability to leverage light has been extensively applied to the quest of imaging systems with increasing resolution. Over the last two centuries, advances in our understanding of how light behaves in an imaging system have propelled the field of optical microscopy from the Abbe diffraction limit and phase contrast imaging to the fields of super-resolution and interferometric scattering microscopy (*1–5*). While near-field, scanning probe techniques such as SNOM and STM have yielded atomic and bond resolved images of the surfaces of materials (*6, 7*), the contact free and relatively simple design of the optical microscope remains the most widespread way of imaging systems, ranging from biological specimens to functional material systems, and is present in almost all experimental labs across the world. Furthermore, a light wave carries with it both spatial and temporal information and access to the temporal axis of an imaging light field could provide a wealth of useful information in optical microscopy.

On the scale of a temporal optical cycle (~1.6 fs at 500 nm), widespread optical microscopes have thus far remained time-integrated detectors, discarding the temporal domain of the light wave through the use of square-law photodetectors in the image plane. This simplification and loss of information is exemplified by the computational design and modelling of optical imaging systems, where a broadband source is set up to radiate optical frequencies to the plane of interest by computationally solving Maxwell's equations on a grid (with techniques ranging from Finite Difference Time Domain (FDTD) to Pseudo-Spectral Spatial Domain (PSSD) (*8*)) and after interacting with the sample, the electric field is then measured by a monitor and propagated to the image plane using Fourier optics (*9*) where its mean squared intensity is computed, entirely losing the temporal information encoded in the electric field by the sample.

Further, while this computational approach has proven immensely successful in understanding light waves in an optical microscope, it fails to describe complex systems that violate the underlying assumptions typical to FDTD and cannot be used as a tool to discover new physical phenomena. For example, modelling a frequency dependent refractive index material or the finite temporal electronic response of medium due to the presence of transient electric fields it is challenging, especially when the underlying timescale of the electronic response is unknown. For instance, a material that undergoes a light induced phase transition or has a delayed polarization response is impossible to study using FDTD, which assumes that the medium is unperturbed by the light field and that the refractive index acts instantaneously on the incoming light field.

It is therefore of observational and informational importance to develop an experimental microscopy platform beyond the square-law detector paradigm, with which we can measure the temporal evolution and formation of images through the full spatio-temporal electric field of light in a conventional widefield optical microscope. While several approaches have been used to both spatially and temporally resolve the electric field in the THz (*10, 11*), a diffraction limit of hundreds of $\mu$m and THz electric-field cycle duration (~ 1ps) are far lower than those that could be obtained with the visible light, standard to the broader microscopy community. Further, measurements that can resolve the electric field of light typically rely on the use of carrier envelope phase (CEP) stable electric fields (*12, 13*), making these experiments even more expensive and challenging, and far from a routine platform with which to study the electric fields in a microscope.

**Results**

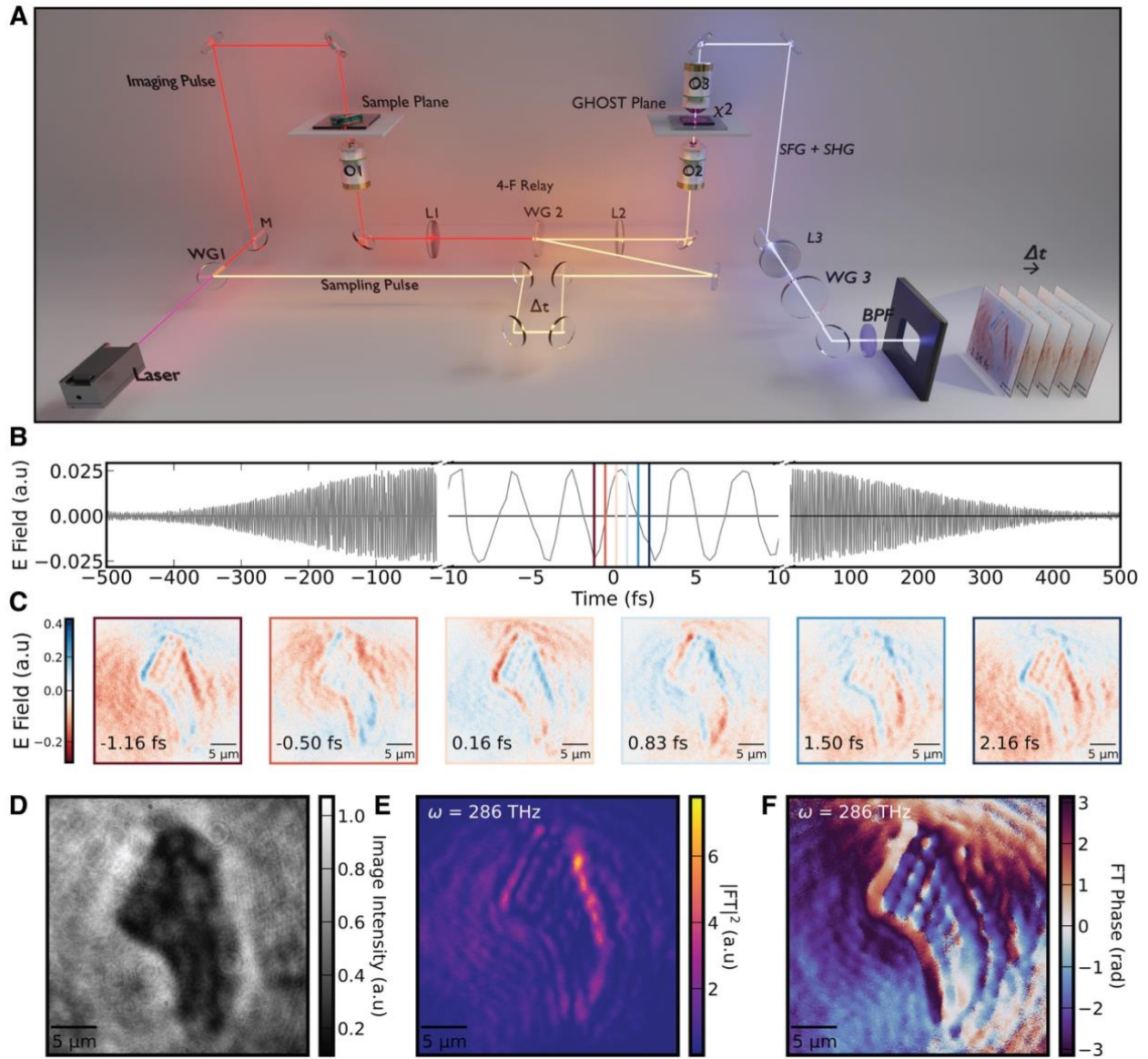

**Fig 1: A.** Schematic of the setup, where a CEP unstable laser pulse is divided with a polarizer (WG1) into two pulses. The weak imaging pulse (red) is loosely focused onto a sample in the focal plane of a high NA oil immersion objective (O1), the transmitted/scattered fields are then collected by the objective and relayed using a 4-F line onto the sample plane of an identical objective (O2), where a $\chi^2$ crystal generates SFG between the imaged electric field and an intense widefield sampling pulse. A third 0.9 NA air objective (O3) images the interference between the SFG and SHG generated onto an emCCD through another 4f line with 255x magnification. As the delay (Δt) between the imaging pulse and sampling pulse is varied, the spatiotemporal electric field is mapped out. **B.** The spatially integrated electric field of a 1030-nm oscillator output, zoomed in between -10 and +10 fs. **C.** Electric field images of the pulse scattering through a flake of $MoTe_2$ at the delays shown in **B** (vertical lines). **D.** Standard absorption image measured on the emCCD of the flake of bulk $MoTe_2$, akin to measuring the square of the electric field in the frequency domain. **E.** Each pixel is Fourier transformed and the spatial FT power is shown at 286 THz with similar features to the standard absorption image, demonstrating the equivalence of the time and frequency domain approaches. **F.** The FT phase image at 286 THz image shows very clear signatures of phase accumulation due to scattering at the flake edges and appears sharper than the power image demonstrating the standard phase contrast imaging mechanism, but in the time domain.

Here we overcome these constraints and develop a highly coherent near-infrared electric-field resolved microscope. In contrast to conventional electro-optic sampling, we utilized a special case (*14*) of the recently developed generalized heterodyne optical sampling technique

(GHOST) (*15*) that allows for measurement of petahertz-scale photons without the need of the CEP-stabilized pulses. We push this technique to its spatio-temporal limit by extending it to a doubly conjugated oil immersion high NA microscope, where our relatively simple experimental platform is able to resolve, for the first time, the formation of images in the plane of a widefield transmission microscope using a CEP-unstable laser light source (*14*).

We demonstrate the unprecedented spatio-temporal information available through this approach by utilizing a long coherence length laser oscillator output, with CEP-unstabilized light pulses centered at 1030 nm wavelength. The pulses are split into two identical orthogonally polarized phase-locked replicas (sampling pulse and imaging pulse) with a wire-grid polarizer WG1. After interacting with the sample, the weak imaging pulse (red line) is collected with a widefield high NA oil immersion objective (O1). The image is conjugated into the identical GHOST plane with a 4-F geometry and a second (identical) high NA oil immersion microscope (O2) with a 5 $\mu$m thick z-cut BBO $\chi^2$ nonlinear crystal present (GHOST plane). In the GHOST plane we utilize a generalized heterodyne optical sampling technique (GHOST) developed in (*15*), where a strong, spatially uniform sampling pulse (derived from the same main laser pulse, yellow), is passed through a delay stage and recombined with the imaging pulse by a wire grid polarizer WG2. This collimated sampling pulse is then focused into the back focal plane of the second microscope (O2), generating a homogenous widefield second harmonic field in the GHOST plane along with a spatially varying sum-frequency field of the sampling and imaging pulses in the BBO crystal plane. Since both, sampling and imaging pulses are identical, the second harmonic and the sum-frequency pulses are of the same wavelength (515 nm) and constitute a SHG/SFG channel for balanced GHOST electric field sampling with CEP-unstabilized pulses (*14*). These fields are then imaged by a third 0.9 NA air objective (O3) at 255x magnification on an emCCD, where they interfere after passing through a polarizer (WG3) in-line to yield an intensity that is proportional to the electric field of the imaging pulse when measured through a bandpass filter (BPF) that isolates the heterodyned SFG – SHG signal, thereby detecting the imaging field. As a delay line between the imaging and the sampling pulse are varied ($\Delta$t), the image on the emCCD maps out the full spatiotemporal electric field in the microscope plane with 100-as temporal resolution and 200-nm spatial resolution. The frequency domain bandwidth of temporal features that we can resolve is given by thrice the bandwidth of the sampling pulse, due to the nature of this SHG – SFG GHOST scheme (see Supplementary Text for details).

To study how images form in the sample plane, we utilize a narrowband, long coherence length laser pulse (oscillator output of our laser system), as both the imaging and sampling pulse (1030 nm ± 10 nm, 200 fs) and study a thick flake of MoTe$_2$ due to the presence of absorptive and refractive features in this spectral region. As shown in Fig. 1B, the measured spatially integrated electric field of the imaging pulse does indeed represent the oscillator output. Zooming into the central 20 fs of the pulse, Fig. 1C shows the full spatial electric field scattering around and transmitting through the MoTe$_2$ flake at 6 delay times (frame colors corresponding to vertical lines in Fig. 1B). We observe the expected effects of the local complex refractive index of the flake (spatial attenuation and retardation), and over a single cycle the spatial electric field returns approximately, *but not exactly*, to its original spatial field. Upon performing a temporal Fourier transform (FT) of each pixel, we can see that features in the standard absorption image (Fig. 1D) are captured by the FT Power at the peak pulse frequency (286 THz), albeit with higher spatial resolution due to the non-linear detection scheme in the $\chi^2$ crystal plane (*16*). We also note that the appearance of a bright edge is likely due to the formation of a waveguided mode in the 2D flake which has been previously reported in van der Waals materials (*17*). The FT Phase at 286 THz (Fig. 1F) however, captures the edges much more clearly than the standard absorption image (Fig. 1D), consistent with the advantages of phase contrast imaging in a system with a large sample-environment refractive index difference (*2*).

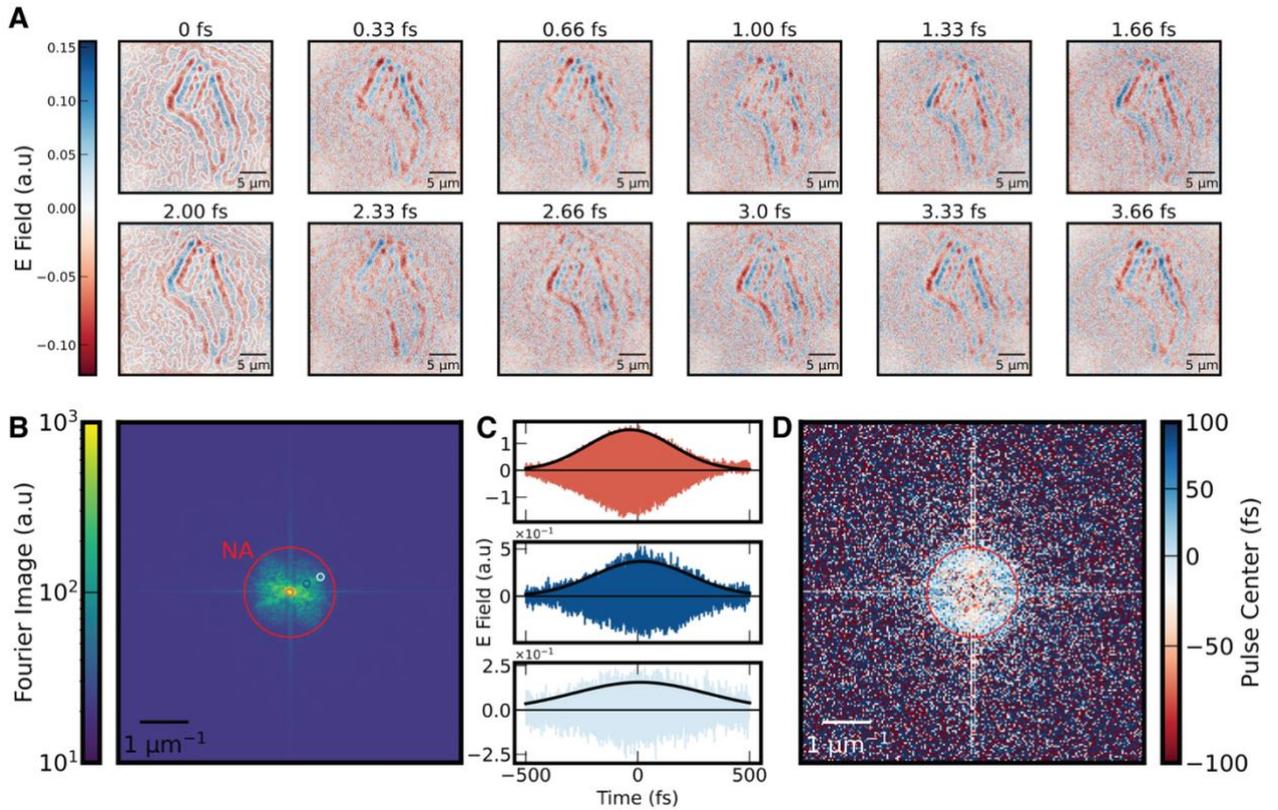

**Fig 2: A.** By suppressing the spatial DC terms in the electric field images in Fig 1 (through a 2D FFT and gaussian suppression near (kx,ky) = (0,0),) slow varying image components are removed, revealing only the temporal formation and interference of the in-plane scattered electric field components around the flake of $MoTe_2$. A standing wave appears to form as demarcated by the zero-field contours in the left two most images (approximately a half optical cycle apart). **B.** A temporally integrated 2D Fourier transformed image, displaying information within the NA of our optical system. **C.** Electric fields of the Fourier image moving away from the central DC component (circles shown in **B**), displaying a clear dependence on |k|. Gaussian fits to the pulse envelope (modified Hilbert transform) are shown in black. **D.** From the Gaussian fits at each pixel in Fourier space, we observe that the high |k| vector components have a markedly different arrival time from the DC part of the image, demonstrating the delayed buildup of in-plane interferometric contrast fringes (see Fig. S1 for a benchmark |k| delay baseline with no sample present).

Upon closer inspection of Fig. 1E and F, we observe local interferometric structures in the FT Power and Phase that are not very clearly present in the standard absorption image (Fig. 1D, which is taken with the exact same effective 12-F setup). These sorts of fringes have been seen in widefield microcopy of a broad set of a materials from batteries to 2D material (*18*), and while the interferometric scattering (iSCAT) mechanism has been used to explain them, full experimental access to their formation dynamics have remained unavailable (*19*). Hence to gain further insight into these spatial fringes, we perform a 2D spatial FFT of the electric field image at every time step. We then suppress the spatial DC components (kx,ky = 0,0) by multiplying each of the 2D FFT images by and inverted 2D Gaussian (sigma = 6 pixels, i.e, 213 $\mu m^{-1}$) centered at (kx,ky = 0,0) with 0 intensity at the center and decaying to a weight 1 radially away, smoothly lowering the weight of the DC components without causing ringing from hard filtering. We then inverse 2D FFT these softly filtered images to retrieve only local spatial distortions in the electric field. This can be equivalently considered as capturing the in-plane interference of the local *scattered* electric field, as if the field did not scatter off the object, there are no local field distortions in the x-y/imaging plane (see Fig. S1 for equivalent experiments and analysis with no sample present).

The temporal slices over 2 optical cycles are presented in Fig. 2A where the emergence of interferometric fringes in the sample plane can be seen. These can be understood as follows – at the interface between air and the MoTe$_2$ flake in the sample plane, there is a sudden large change in the real and imaginary refractive index. This causes scattering (i.e., local refraction and reflections) into the sample plane, where a component of the electric field propagates. Every flake edge acts as a source of these in-plane scattered wavefronts, which propagate and interfere in the sample plane. As the zero-field contour lines in the two left-most panels (0 and 2 fs, approximately a half optical cycle apart) demonstrate, the positive regions all turn negative and vice versa, demonstrating the transient formation of standing waves in the sample plane. We note that it is due to our use of a widefield imaging modality and a high NA oil immersion system that we can capture the high angle scattered waves.

Looking at the time averaged 2D |FFT| map in Fig. 2B, we can also see the generation of a clear line in the spatial frequency domain due to the formation of these standing waves. By isolating the electric field of different |k| values along this line of maximum scattering (circles in Fig. 2B), we can see that as we go to larger |k| values, away from the spatially uniform 'DC' image, the electric field envelope position changes substantially (Fig. 2C), i.e., on average the electric fields at higher in-plane k-vectors appear to form later than the main plane wave (DC part). By fitting the pulse envelope (extracted using a modified temporal Hilbert transform) to a Gaussian (black lines Fig. 2C), we can extract the pulse delay as a function of (kx,ky), demonstrating the delayed formation of interferometric fringes as a function of |k| in Fourier space (Fig. 2D). We have repeated this on several different MoTe$_2$ flakes (of varying thickness) and in FDTD simulations made to emulate our flakes results and find it to be consistent across them (see Fig. S2-4). The delayed formation of interferometric fringes can be understood by considering the finite time the scattered waves take to propagate in-plane from the edges of the flake and interfere. This naturally scales with the speed of light over the flake dimensions, with the precise structure of the group delay as a function of (kx,ky) being a measure of the in-plane dispersion of the photonic modes in the flake. This delayed interference effect could have implications for improved contrast background free, time gated scattering microscopy (see comparison to time averaged and standard microscopy in Fig. S7). Further, the implications of the buildup of in-plane standing waves on the electronics of the underlying material could provide routes to strong light-matter coupling, for example in systems with engineered defects (*20*).

To bolster these ideas, we perform the same analysis in real space, i.e., at each point in space we use a temporal Hilbert transform we extract the pulse envelope and by fitting it to a Gaussian we can simplify the analysis to three parameters: the local field magnitude, pulse delay and width. As seen in Fig. 3A, the Gaussian fitting extracts the same field amplitude as the above analysis, and the pulse delay map (Fig. 3B) demonstrates that of the interferometric, oscillatory features appear at a different time to the background DC component of the field images. Strangely however, we also observe that the pulse width (Fig. 3C) is not uniform in space and there are regions, especially on the interior of the flake where the signal is high, and the pulse is broadened (see Fig. S8 for a spatial map of fit errors, below our observational threshold). With out current signal-to-noise ratio we are unable to easily clarify the origin of this effect as either a non-linear absorptive loss of spectral content or as the chirping of the input pulse.

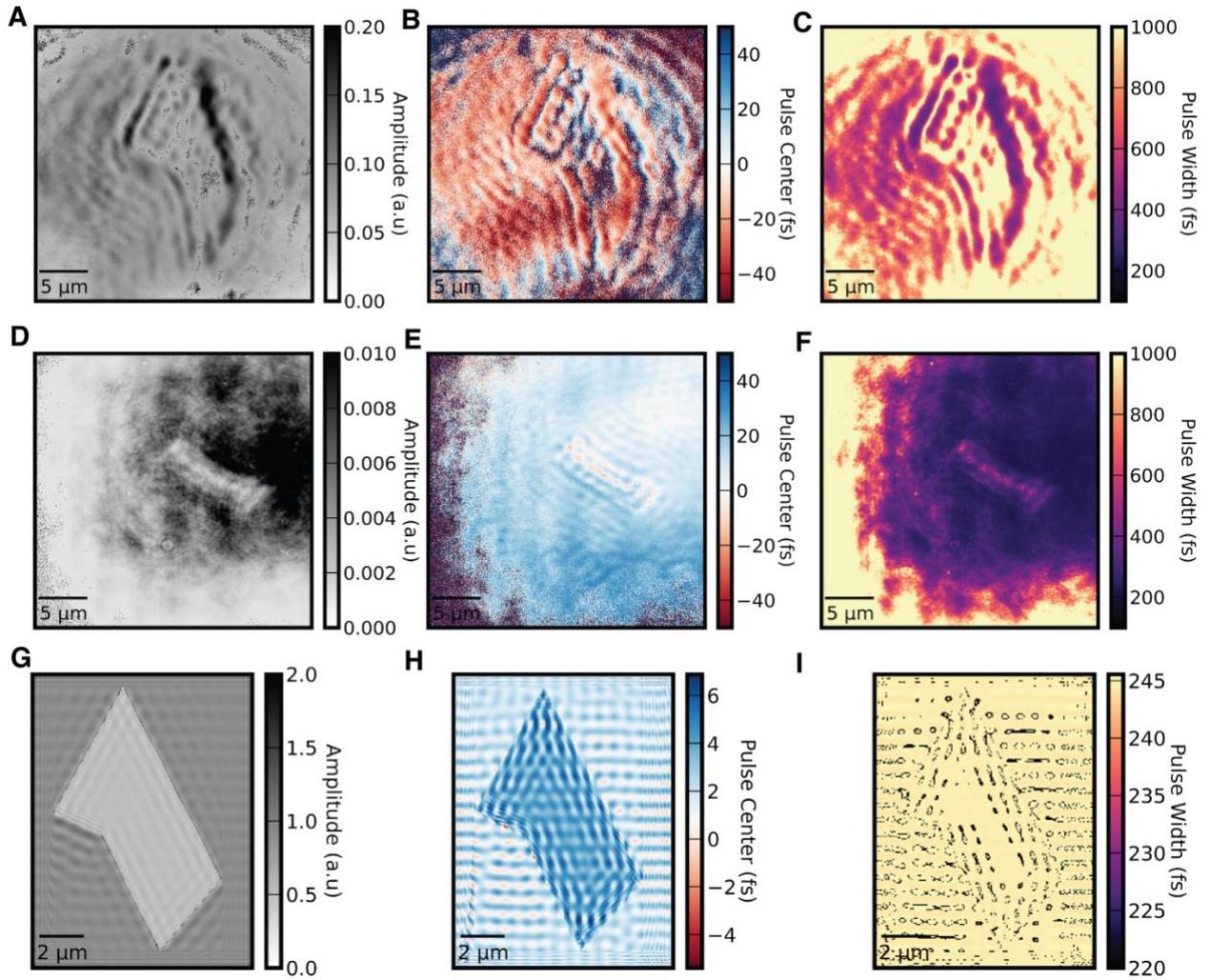

**Fig 3: A.** Moving back to real space and fitting the Gaussian envelope of the 200-fs pulse, we find that indeed the interferometric fringes appear later (**B.**), along with the intriguing observation that the pulse width is spatially dependent (**C.**). **D., E.** and **F.** demonstrate this observation on an entirely different flake of MoTe$_2$. **G., H.** and **I.** represents the same analysis on a FDTD simulation of a similarly shaped and sized flake of MoTe$_2$, consistent with the delayed appearance of interferometric fringes but with no pulse width dependence in space at all.

These observations are consistent across different flakes, with a second example shown in Fig. 3 D., E. and F. In Fig. 3 G., H. and I. we perform FDTD simulations on a spatial n,k material with the same refractive index and dimension of our MoTe$_2$ flake in Fig. 3A, and with the same pulse characteristics as our experiment. We note that inherent to the FDTD simulation is the assumption that the material refractive index is static in time and not influenced by the incident wave. We find that we can reproduce the delayed formation of the interferometric fringes (see Fig. S3 and S4 for polarization dependence of build-up of interferometric scattering signals, and Fig. S5 and S6 for analogous FDTD simulations and analysis) that we observe in the experiment. However, the pulse broadening effects are not reproduced by the FDTD simulation. Further, as the flake is micron scale in size and the pulse is already spectrally narrow (200-fs, transform limit), this cannot be explained as a refractive index/GVD dispersion of the pulse (that is not captured in our FDTD simulations), leaving only electronic effects as a plausible explanation. We hypothesize that the local standing waves seen in Fig. 2 generate large polarization fields in the plane of the material, which decays in finite time, broadening the pulse where there is the largest standing wave field. This observation is critically not captured by FDTD simulations, demonstrating the advantage of our experimental approach in establishing the ground truth of what happens in the sample plane of an optical microscope when imaging a material.

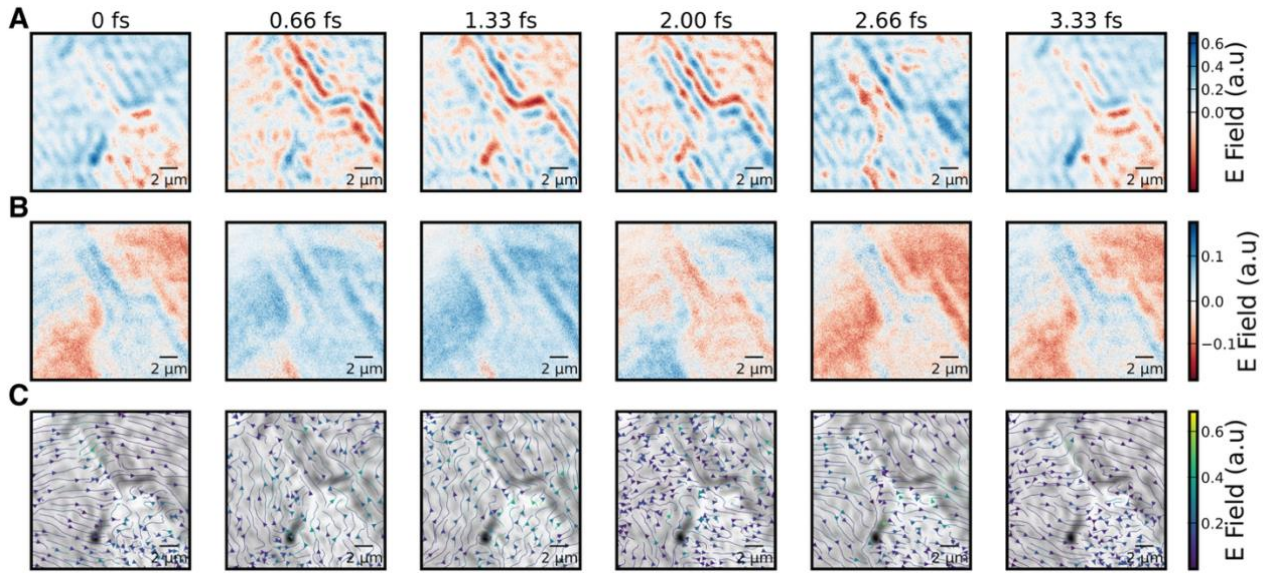

**Fig 4: A.** Time series of a flake of MoTe$_2$ with the $\chi^2$ crystal in the 2$^{nd}$ image plane positioned to phase match one polarization of the scattered field. **B.** Time series of a flake of MoTe$_2$ with the $\chi^2$ crystal in the 2$^{nd}$ image plane positioned to phase match a different (not completely orthogonal) polarization of the scattered field. **C.** By aligning the measurements in **A.** and **B.** approximately in time and space, at each spatiotemporal point we have access to the full in-plane vector field of the optical pulse in the microscope plane. From this we plot the field lines and can watch them evolve with time.

To demonstrate the power of our experimental approach, we additionally image the full vector electric field lines in the microscope plane through a sample as a function of time. To obtain this information, we utilize the unique property of the z-cut bbo crystal that is used as a GHOST medium. In our special balanced case, the medium produces a second harmonic from the strong sampling pulses as well as sum-frequency generation between the strong sampling pulse and the weak imaging pulse that contains information about the scattering dynamics in the sample. The trigonal symmetry of the z-cut BBO allows us to choose polarizations of the SHG and SFG light. By placing a wire grid polarizer after the BBO crystal and rotating the crystal we effectively choose the polarization of the imaging pulse electric field of which is being sampled in time and space. Fig. 4A and 4B are time series images of the flake recorded with two angles (30 degrees apart) of the BBO crystal. By taking these two measurements and approximately spatially and temporally overlapping them, we can measure the evolution of electric field lines in the material as a function of time (Fig. 4C.). Since the photon energy (~ 1.2 eV) utilized in our experiment is substantially smaller than the work function of the MoTe$_2$ flake (~ 4.5 eV), the charge must be net neutral in the material, implying that we can additionally enforce that all field lines are continuous in our material as a function of time. The power of this vector field imaging modality can be seen in Fig. 4C. where we are able to observe how electric field lines propagate around the edge of the material and which sections of the flake have the strongest induced optical dipoles.

To summarize, we have developed an experimental scheme that is capable of spatiotemporally resolving CEP-unstable electric fields of light in the plane of a widefield transmission optical microscope. Using this we have unveiled, for the first time, experimental insight into the precise dynamics of image formation; from the delayed onset of interferometric features due to in-plane scattered standing waves, to pulse broadening dynamics beyond linear optical effects typically captured by FDTD. Finally, we showcase the broad applicability of our experimental scheme, by resolving the in-plane electric vector field lines during photoexcitation. In this paper we have focused on utilizing the laser oscillator output as our imaging wave pool, which gives us narrow bandwidth and long coherence time/length optical pulses, but with the downside of the SFG-SHG GHOST frequency detection window being narrow. We note that modification of this setup is possible to compressively sense a 5-fs 200-THz bandwidth optical pulse which we have additionally achieved and this will be the subject of another paper. Our work allows experimental access to the formation of the full broadband, vector spatio-temporal light field in the ubiquitous transmission optical microscope, pushing our understanding and utilization of centuries-old optical microscopy to its ultimate limits, with broad applications ranging from improved benchmarks for computational microscopy techniques to laying the foundation for entirely new electric field imaging modalities for material systems.

**Acknowledgements:** We thank Ryo Mizuta Graphics for help with the Blender images of the setup.

**Funding**: A.A. acknowledges funding from a Trinity College Title A Fellowship. D.A.Z. acknowledges funding from the Swiss National Science Foundation (grant no. PZ00-2_232911).

**Competing interests**: The authors declare that they have no competing interests.

**Data and materials availability**: The data that support the plots within this paper and other findings of this study are available at the University of Cambridge Repository (https://doi.org/XXXXX).

**Rights Retention Statement:** This work was funded the UKRI. For the purpose of open access, the author has applied a Creative Commons Attribution (CC BY) license to any Author Accepted Manuscript version arising.